\documentclass[manuscript,screen]{acmart}
\usepackage{array}
\usepackage{array,multirow,graphicx}

\AtBeginDocument{%
  \providecommand\BibTeX{{%
    \normalfont B\kern-0.5em{\scshape i\kern-0.25em b}\kern-0.8em\TeX}}}

\copyrightyear{2022} 
\acmYear{2022} 
\setcopyright{rightsretained} 
\acmConference[CHI '22 Extended Abstracts]{CHI Conference on Human Factors in Computing Systems Extended Abstracts}{April 29-May 5, 2022}{New Orleans, LA, USA}
\acmBooktitle{CHI Conference on Human Factors in Computing Systems Extended Abstracts (CHI '22 Extended Abstracts), April 29-May 5, 2022, New Orleans, LA, USA}
\acmDOI{10.1145/3491101.3503744}
\acmISBN{978-1-4503-9156-6/22/04}

\begin{document}

\title{Empathy-Centric Design At Scale}

\author{Andrea Mauri}
\email{a.mauri@tudelft.nl}
\affiliation{%
  \institution{Delft University of Technology}
  \streetaddress{Landbergstraat, 15}
  \city{Delft}
  \country{Netherlands}
  \postcode{2628 CE}
}
\author{Yen-Chia Hsu}
\email{y.hsu-1@tudelft.nl}
\email{x}
\affiliation{%
  \institution{Delft University of Technology}
  \streetaddress{Landbergstraat, 15}
  \city{Delft}
  \country{Netherlands}
  \postcode{2628 CE}
}

\author{Marco Brambilla}
\affiliation{%
  \institution{Politecnico di Milano}
  \streetaddress{Piazza Leonardo da Vinci, 32}
  \city{Milan}
  \country{Italy}}
\email{marco.brambilla@polimi.it}

\author{Aisling Ann O'Kane}
\email{a.okane@bristol.ac.uk>}
\affiliation{%
  \institution{University of Bristol}
  \streetaddress{Senate House, Tyndall Avenue}
  \city{Bristol}
  \country{United Kingdom}
  \postcode{BS8 1TH}
}

\author{Ting-Hao `Kenneth' Huang}
\email{txh710@psu.edu}
\affiliation{%
  \institution{Pennsylvania State University}
  \streetaddress{201 Old Main, University Park}
  \city{State College}
  \country{USA}
  \postcode{16802}
}

\author{Himanshu Verma}
\email{h.verma@tudelft.nl}
\affiliation{%
  \institution{Delft University of Technology}
  \streetaddress{Landbergstraat, 15}
  \city{Delft}
  \country{Netherlands}
  \postcode{2628 CE}
}

\renewcommand{\shortauthors}{Mauri, et al.}

\begin{abstract}
  %EmpathiCH aims to bring together different expertise to investigate how digital tools can be used to elicit empathy between multiple stakeholders at scale.
  %The motivation behind this works lies in the necessity of new digital-supported methodologies to design at-scale solutions to societal problems that are balanced, inclusive, and aware of their effect on society.
  %This full-day workshop includes participants' presentations, an inspiring keynote talk, and a panel of experts.
  %Stimulated by this discussion, the expertise, and experiences of the participants will be combined during a participatory session to shape the future of the ``Empathy Design at Scale'' research agenda.
EmpathiCH aims at bringing together and blend different expertise to develop new research agenda in the context of ``Empathy-Centric Design at Scale''.
The main research question is to investigate how new technologies can contribute to the elicitation of empathy across and within multiple stakeholders at scale; and how empathy can be used to design solutions to societal problems that are not only effective but also balanced, inclusive, and aware of their effect on society.
Through presentations, participatory sessions, and a living experiment---where data about the peoples' interactions is collected throughout the event---we aim to make this workshop the ideal venue to foster collaboration, build networks, and shape the future direction of ``Empathy-Centric Design at Scale''.
\end{abstract}





\maketitle

\section{Background}

%\textbf{Andrea: the subsection headings may be removed in the end, now they are there to set up the structure}

When designing solutions at a large temporal and geographical scale to address societal problems, it is essential to consider both human needs (e.g., safety) and values (e.g., inclusiveness, ethics)~\cite{Demirel2021}.
%When designing at scale to address societal problems it is important to consider both human needs (e.g., safety) and values (e.g., inclusiveness, ethics)~\cite{Demirel2021}.
%
The advance of new technologies (e.g., AI, novel sense-making algorithms) and the increased affordability of sensors enable designers to collect data from a large and diverse set of users.
%The advance of new technologies such as AI, novel sense-making algorithms and the increase affordability of sensors, allows designers to access data about a large and diverse set of users.
%
The information in the user-contributed data is primarily used to design solutions through optimizing quantitative metrics, such as efficiency~\cite{Lee2015}, environmental and economic indicators~\cite{Simon2017}, or criminality~\cite{toole2011spatiotemporal,mohler2011self}.
%However, this information is often used to design a solution by optimizing quantitative metrics such as: efficiency~\cite{Lee2015}, environmental and economic indicators~\cite{Simon2017}, or criminality~\cite{toole2011spatiotemporal,mohler2011self}.
%
However, this optimization approach can lead to solutions that favor the decision-makers rather than balancing the needs of all stakeholders, which may result in not inclusive solutions and a lack of trust among the individuals~\cite{clarke2021socio}.
For instance, a policy that is developed to maximize public safety with criminality indicators (such as~\cite{mohler2011self}) will advocate a mass deployment of security cameras, which can have privacy concerns and may give people the uncomfortable impression of surveillance.
Another example is that an urban intervention, designed using optimization approaches, can lead to inappropriate solutions that systematically deny services for part of the population\footnote{Unpleasant Design \& Hostile Urban Architecture: \url{https://99percentinvisible.org/episode/unpleasant-design-hostile-urban-architecture/}}.
%For instance, a security policy built to maximize people safety - using as metric indicators such as~\cite{mohler2011self} - will advocate a mass deployment of security cameras - with issues related to privacy and giving the impression to people of being under surveillance; or, a similarly designed urban intervention will lead to solutions aimed at excluding part of the population\footnote{Unpleasant Design \& Hostile Urban Architecture: \url{https://99percentinvisible.org/episode/unpleasant-design-hostile-urban-architecture/}}. 
%
While these solutions may successfully address the main problems of concern, in the long term, their unseen effects can have a negative social impact on a part of the population, typically minorities or vulnerable individuals~\cite{barocas2016big}.
%While these solutions may successfully address the main problems they were designed for, in the long term their unseen effects will have negative consequences on part of the population, typically minorities or vulnerable individuals~\cite{barocas2016big}.

To take social impact into account, we need to develop human-centered design approaches that \textbf{works at a large scale}, is data-driven, and can fully capture the diverse and complex landscape of people’s needs, interests, and values (such as those described in~\cite{davis2015value}).
%We believe this phenomena is caused by the inability of current data-driven human-centered design approaches to fully capture the diverse and complex landscape of people’s needs, interests and values~\cite{davis2015value}.
%
%At the best of our knowledge, this research field is under-explored and requires more attention~\textcolor{red}{[citation needed]}.
%
Current participatory design methods (such as Community Citizen Science~\cite{hsu2020human}, workshops, and focus group interviews) have the benefits of centering on people's perspectives~\cite{steen2011benefits}.
%Participatory design methods - such as Community Citizen Science~\cite{hsu2020human}, co-design and co-creation workshops, and focus groups - have the benefits of focusing more on people's perspectives~\cite{steen2011benefits}.
%
However, these methods can suffer from scalability and hyper-locality problems~\cite{carroll2015reviving}.
This means that the resulting solutions are designed for a specific local context, which can be difficult to generalize to broader contexts.
%However, they suffer of scalability and hyper-locality problems - i.e., it is difficult to apply the findings to design solution for wider context; also, they do not provide the designers with the tools to integrate conflicting perspectives.
%
Moreover, these methods rely on designers' experiences in integrating conflicting perspectives at a small scale~\cite{klein1989conflict}, and there is a lack of digital tools that can support designers in handling conflicts at a large scale.
%Also, they do not provide the designers with the tools to integrate conflicting perspectives.

Besides the scalability concerns, we further argue that it is crucial to develop co-design methods and tools to elicit \textbf{empathy} among stakeholders to establish trust and form sustainable relationships.
%We believe that developing \textbf{empathy} between the stakeholders is a key factor to address this problem.
%
We believe there is a need for novel digital-supported methodologies to support the elicitation and integration of human needs in human-centered design approaches through the development of empathetic relations.
Current design research considers empathy a crucial factor in understanding better about people's needs~\cite{segal1997empathic}.
%In design, empathy is considered a crucial aspect to be utilized to get a better understanding of people needs~\cite{segal1997empathic}.
%
Empathy is defined as ``the intuitive ability to identify with other people’s thoughts and feelings – their motivations, emotional and mental models, values, priorities, preferences, and inner conflicts''~\cite{mcdonagh2006empathic}, which means going beyond ``knowing the user'' and understanding how it \textit{feels like} to be that person~\cite{Merlijn2009,wrigth2008}.
%Defined as "the intuitive ability to identify with other people’s thoughts and feelings – their motivations, emotional and mental models, values, priorities, preferences, and inner conflicts"~\cite{mcdonagh2006empathic}, it means going beyond "knowing the user" and it involves understanding how it \textit{feels like} to be that person~\cite{Merlijn2009,wrigth2008}.
%
For decades, scholars have studied how forming empathetic relationships between designers and users results in better products or services~\cite{wrigth2008}.
%To this end, different scholars have been studied for decades how building empathetic relations between designers and the users results in better products or services~\cite{wrigth2008}.
%
A large variety of methodologies have been developed to get the designers ``into the shoes'' of the people they are designing for: cultural probes~\cite{mattelmaki2002empathy}, focus groups~\cite{Froukje2005,mcdonagh2000using,mikko2021}, storytelling~\cite{visser2007,postma2012,Mieke2014}, and simulation of the user's condition~\cite{farmer2014change,Bertrand2018,bennet2019}.
%\textcolor{red}{However, current HCI research in empathy [XXX].}

HCI and many other disciplines have investigated the importance of empathy: in patients medic relation~\cite{milcent2021using}, education~\cite{whitford2019empathy,BACHEN201677}, racial bias reduction~\cite{patane2020exploring}, gaming~\cite{gilbert2019assassin,BACHEN201677}, design~\cite{segal1997empathic,gasparini2015perspective,yuan2014empathy}; and with different technologies like, virtual reality~\cite{ventura2020virtual,milcent2021using,BACHEN201677}, mobile~\cite{o2014gaining} and wearable devices~\cite{rojas2020,hassib2017}, and artificial intelligence~\cite{Sharma2021}.
Building on previous work, \textbf{in this workshop, we take a multi-stakeholder perspective} to create empathetic links both across and within various groups of people, including designers, decision-makers, scientists, and citizens.
%Building on this body of literature, \textbf{in this workshop we now take a multi-stakeholders perspective}, where not only it is important to create empathetic links between designers and users but also between different individuals of the population.
%
%The importance of empathy has been investigated in different domains and technologies, including patients medic relation~\cite{milcent2021using}, education~\cite{whitford2019empathy,BACHEN201677}, racial bias reduction~\cite{patane2020exploring}, gaming~\cite{gilbert2019assassin,BACHEN201677}, design~\cite{segal1997empathic,gasparini2015perspective,yuan2014empathy}, virtual reality~\cite{ventura2020virtual,milcent2021using,BACHEN201677}, mobile~\cite{o2014gaining} and wearable devices~\cite{rojas2020,hassib2017}, and artificial intelligence~\cite{Sharma2021}.
%

This workshop aims to bring together different scholars---including computer scientists, social scientists, designers, psychologists, policy-makers, and practitioners from other disciplines---to share their knowledge, experience, and ideas about working with empathy for large-scale societal impact.
Our ambition is to become a venue that blends the aforementioned expertise to define future research and collaborations in the \textbf{``Empathy-Centric Design at Scale''} area.
%The aim of this workshop is to bring together different scholars - computer scientist, social scientist, designers, psychologists, policy-makers, and practitioners from other disciplines - to share their knowledge, experience and ideas about working with empathy, with the ambition to become a venue where we can blend together the expertise mentioned above to define future research and collaborations.
This diversity of expertise is fundamental since we want to investigate how new technologies can support the design \textit{with} and \textit{for} empathy in the following cases:
\begin{itemize}
    \item \textbf{\textit{For different purposes.}}
    Solutions to societal problems may take many forms, such as a policy, a service, or a product.
    Depending on the context, the required methods and the supporting technologies in the design processes of these solutions can be very diverse.
    %The characteristics of the design processes needed in these different cases are very diverse, together with the technologies that can be applied.
    %
    Thus, we need to identify and evaluate the roles of empathy in co-designing these solutions.
    Which facets of empathy (cognitive or emotional) can be elicited in which context?
    How can we embed these empathy facets in different co-design contexts that lead to inclusive solutions?
    How can we develop and use new technologies (e.g., virtual reality, AI, internet of things, and wearable sensors) to facilitate empathy design for various purposes?
    %For this reason, we need to identify how empathy can play a role in this case, how it can lead in more inclusive solutions and how new technologies can help.
    %Example of questions are: how new technologies (e.g., virtual reality, AI, IoT/wearable sensors) can be used to embed empathy in the design process of such different cases?
    %Which facets of empathy (cognitive or emotional) can be elicited and how?
    %
    \item \textbf{\textit{Across and within multiple stakeholders.}}
    Societal problems affect a large variety of stakeholders, each with a different set of needs and values.
    In order to design inclusive solutions, it is important not only for the designers and decision-makers to understand the citizens but also for the people belonging to the same group to understand each other's perspectives.
    %In order to design an inclusive solution, not only is important for the designer to understand the user, but also it is crucial for the people belonging to the group of users to understand each other different perspectives.
    %
    How can new technologies empower creating empathy across and within multiple stakeholder groups?
    %So, how new technologies can empower empathy between the different stakeholders?
    %
    What are the co-design methods that can make different and diverse communities empathize with each other?
    %How can we make different and diverse communities empathize with each other?
    %
    How can empathy be employed to reach broader communities and gather unbiased data?
    %How empathy can be exploited to gather more unbiased data?
    \item \textbf{\textit{At different scales.}}
    Solutions can affect society at different scales, such as a policy at the national level, an urban intervention at the city level, or a plan within an organization.
    %Solutions can affect society at different scales: it can be at national level, a urban intervention at city level, or a policy within an organization.
    %
    Moreover, different scales can have diverse sets of technological and societal challenges.
    %Beside the magnitude, different cases come with a diverse set of technological and societal challenges.
    %
    For instance, as the scale becomes larger, reaching out to diverse stakeholder groups (especially underserved communities) and representing different perspectives can be much more challenging.
    %For instance, bigger the scale, more a designed solution have to take into the account the diversity of the population.
    %
    How can we develop or adopt digital tools to elicit empathy at different scales, especially at the large scale that is hard to tackle using tools that were developed to support small-scale design activities?
    %So, questions related to this case may be related to the \textbf{scalability} and \textbf{interoperability} of the new technologies - for example, to what extent a technology developed to elicit empathy in one case be applied to another? 
\end{itemize}

We want to take the opportunity of this workshop to study the role of empathy in a design process and to provide scholars with data and insight to kick-start future research on this topic.
To do so, this workshop will be a living experiment, where all the interactions---onsite and remote---between the participant will be recorded (e.g., talks, questions, comments), annotated, combined with sensor data from wearable devices, and made public for everyone to access.
We envision the outcome from the discussions---e.g., action points for a common research agenda---together with the collected dataset to be the starting point where researchers from many disciplines can work together and contribute to the area of Empathy-Centric Design at Scale.

%[copied from a paper]"These implementation strategies and effects of these proposals must also be creatively and critically investigated. This pertains to societal institutions, such as schools and institutions of higher education; legal institutions; local, national and international policy makers; and the very practices of designers and researchers."

%\subsection{Themes}

%\textbf{Andrea: listing themes in random order, needs to be grouped and categorized}

%\begin{itemize}
    %\item Empathy for the design at different scale: national, city, company
    %\item Empathy for the design of services, products and policy
    %\item Empathy between users, between designers, between user and designers
    %\item Empathy-centric data driven design processes
    %\item Empathy-centric citizen participation, empathy to build communities
    %\item Empathy-empowering (new) technologies. Data mining tools for empathy stuff
    %\item Role of empathy in decision making
%\end{itemize}

%\textbf{Random stuff:}

%Empathy is better w.r.t Adversarial interaction

%Empathy in working environment

%Empathy in review process

%Empathy in question - answer in presentation/talk
\section{Organizers}

Below is a list of the organizers' short biographies.
Their expertise and interests are diverse but in line with the workshop topic, which reflects our goal to have interdisciplinary perspectives and discussions.
They come from different academic fields and industry, and can bring on the table not only a richer set of perspectives but also a diverse and complementary network of people that may be interested in the workshop.

\textbf{Andrea Mauri} (main contact) is a Postdoctoral Researcher at the Faculty of Industrial Design Engineering at Technische Universiteit Delft (Netherlands). He is also a Research Fellow at the Amsterdam Institute for Advanced Metropolitan Solutions. He has a background in applied machine learning and data science. He is interested in the design, implementation, and evaluation of novel computational methods and tools - focusing on hybrid human-AI methodologies - to support the design processes addressing societal problems by integrating human and societal needs and values. (Website: \url{https://www.andreamauri.com})

\textbf{Yen-Chia Hsu} is a Postdoctoral Researcher at the Faculty of Industrial Design Engineering, Technische Universiteit Delft (Netherlands). He studies methods to co-design, implement, deploy, and evaluate interactive AI systems that empower communities. He applies crowdsourcing, data visualization, machine learning, computer vision, and data science to engage and assist communities in addressing local environmental and social concerns. (Website: \url{http://yenchiah.me})

\textbf{Marco Brambilla}  is a full professor at Politecnico di Milano. He manages several research and industrial innovation projects. His research interests include data science, software modeling languages, crowdsourcing, social media monitoring, data-driven innovation, and big data analysis. He is the main author of the OMG standard IFML. He is associate editor of the journals: Web Engineering, Digital, and Advances in Human-Computer Interactions.  (Website: \url{https://marco-brambilla.com})

\textbf{Ting-Hao `Kenneth' Huang} is a tenure-track Assistant Professor at the College of Information Sciences and Technology of the Pennsylvania State University. His research lies in the intersection of Artificial Intelligence and Human-Computer Interaction, imagining new possibilities of human-AI collaborations. Dr. Huang explores the creative and complex domains, such as open conversation, writing support, and automatic storytelling, which seem exceptionally challenging to automate. His work aims to move automation beyond low-level, mundane tasks to augment human creativity and sociability. (Website: \url{https://crowd.ist.psu.edu/index.html})

\textbf{Aisling Ann O'Kane} is a Senior Lecturer in Human-Computer Interaction for Health at the University of Bristol and is Deputy Director of the EPSRC CDT in Digital Health and Care. She has over ten years of experience studying the real world use of health and care technologies outside of clinical settings. (Website: \url{https://research-information.bris.ac.uk/en/persons/aisling-a-okane})

\textbf{Himanshu Verma} is a Tenure-Track Assistant Professor at the Faculty of Industrial Design and Engineering at Technische Universiteit Delft (Netherlands). He has a background in HCI, UbiComp and Social Cognition. He is interested in examining collaboration at scale, and his current research is focused on sensing and modeling of interpersonal collaborative processes and how they can be better supported through wearables. In addition, he is also interested in studying the perceptual, cognitive and experiential aspects of human-AI collaboration. (Website: \url{https://vermahimanshu.com/})
\section{Link to website}
The website will be hosted at a public GitHub repository using the GitHub Pages service\footnote{GitHub Pages: \url{https://pages.github.com}}. 
The website's content will contain the information of organizers, important dates (e.g., submission deadline, notification of acceptance, etc), workshop schedule, updates about the workshop, call of papers text, and the link to submit the manuscript.
We will also make the accepted papers available on the website before starting the workshop.
After the workshop, we will also update the website with the workshop's summary,  output, and results.
\section{Pre-Workshop Plans}

Our goal is to hold an interdisciplinary workshop, including industry and academic researchers from the areas of ACM SIGCHI (e.g., CHI, IUI, DIS, CSCW, UbiComp), web science (e.g., WWW), social science, psychology, artificial intelligence, health, and policy-making.
The organizers are active in these research areas and plan to encourage colleagues and students in their networks to participate in this workshop.

We will distribute the call for papers information through the ACM SIGCHI mailing list, the website that we will set up, and the organizers' professional networks, such as institution mailing lists and social media (e.g., Twitter and Facebook).
We plan to host around 15 to 20 participants in the workshop, which we believe is a suitable size for building a community, networking with each other, and engaging in conversations.

For paper selection and reviewing, the workshop organizers will reach out to more researchers to form a program committee.
We aim to have a good balance of diverse perspectives and topics that are related to the workshop themes.
Each paper submission will receive at least two reviews from the program committee to assess the novelty, provocativeness, quality, and relevance.
Those with well-presented and insightful contributions will be selected.

Before the workshop, we will make the accepted papers and workshop schedule publicly available on our website.
For the accepted papers, we will request slides for a short live presentation and also an 8-minute video from the authors, and we will upload these materials to our website before the workshop.
One week before the workshop, we will ask the participants to familiarize themselves with the papers and the videos.
We will also distribute necessary information (such as a survey to sensitize people) regarding the interactive workshop activities that we plan to engage participants.

During the workshop, we plan to run an experiment related to how empathy affects conversations in academic settings.
From the live experiment, we plan to collect research data to gain insights into the role of empathy in the co-design process.
Information about the experiment is described in the next section.
Before the workshop, we will ask the participants (via email) if they would like to bring their own sensors or device for the data collection process.
We will also distribute an online informed consent form (via email) about data collection to the participants.
\section{Workshop Structure}

\begin{table}[t]
\centering
\caption{Proposed workshop schedule.}
\begin{tabular}{c p{1.5cm} p{11cm}} 
 & Duration & Activity \\
 \toprule
 & 10 minutes & \textbf{Set Up:} login to the Zoom platform (for remote participants only) and greet all people. Eventual sensor check and start of the workshop-long data collection process. \\
 & 15 minutes & \textbf{Welcome:} introduce organizers, participants, workshop objectives and schedule.  \\
  \parbox[t]{2mm}{\multirow{4}{*}{\rotatebox[origin=c]{90}{\textbf{Presentations}}}}& 30 minutes & \textbf{Warm up:} interactive discussion to sensitize the participants on the topic of empathy  \\
& 45 minutes & \textbf{Keynote:} presentation by an invited expert with Q\&A and discussions \\
 & 15 minutes & Short break \\
 & 45 minutes & \textbf{Minute Madness:} participants present their papers in a minute madness style, followed by a moderated discussion session. Presentations should be prepared in advance in Google Slides or as prerecorded video presentations \\
 & 30 minutes & \textbf{Panel:} discussion among experts and participants about how empathy may be applied in the design process of projects or tools \\
 \midrule
 & 30 minutes & Lunch break and social gathering \\
 \midrule
 & 30 minutes & \textbf{Eliciting Themes:} rapid group discussion to elicit unaddressed questions raised during previous workshop activities \\
 & 10 minutes & \textbf{Cluster Themes:} participants and organizers group and categorize the themes in topics of interests\\
 \multirow{10}{*}{\rotatebox[origin=c]{90}{\textbf{Ideation and Mapping Session}}}& 30 minutes & \textbf{Ideation Session:} small groups within Zoom break-out rooms will brainstorm about one topic of interest, and add their thoughts on a shared Miro board \\
 & 15 minutes & Short break \\
 & 30 minutes & \textbf{Group Feedback:} groups present the results of their ideation to all participants for feedback \\
 & 30 minutes & \textbf{Mapping Session:} small groups within break-out rooms and using Miro boards will consolidate their ideas into developed research agendas \\
 & 30 minutes & \textbf{Summarize Discussions:} each group prepares a short presentation to summarize their developed research agendas and how to take them forward \\
 & 15 minutes & Short break \\
 & 30 minutes & \textbf{Group Presentations:} each group presents and discusses the results of ideation and mapping session to all other participants \\
 & 30 minutes & \textbf{Wrap Up:} summarize the workshop, actions on follow-up activities, and take group photos (both onsite and remote). Closing of the data collection process. \\
 \midrule
 & - & Drinks and networking \\
 \bottomrule
\end{tabular}
\label{table:schedule}
\end{table}

%\begin{itemize}
%    \item 1 day workshop with a table structure that contains time and activities
%    \item first half of the day interactive session + data collection (cocteau)
%    \item second half of the day presentations
%    \item we could also run a COCTEAU~\cite{tocchetti2021} experiment during the workshop as a empathy-driven data collection method; it kinda blends in with the narrative of the workshop, it makes it more interactive, it allows us to have concrete future activity, and provides data for future paper.
%\end{itemize}

We propose a full-day workshop with submissions that includes position, work-in-progress, provocation, demo, or poster papers (between 4-6 pages, excluding references).
We are interested in a wide range of novel concepts and perspectives.
The workshop will be held in hybrid form, both in presence and on Zoom\footnote{Zoom: \url{https://zoom.us/}}.
Details of the hybrid setup is mentioned later in Section~\ref{section:hybrid-plan}.
We will use Miro\footnote{Miro: \url{https://miro.com/}} as a way to allow collaborative activities with remote participants.
The entire event is estimated to be around 8 hours with different activities, including presentations, social events, breaks, and discussions as shown in Table~\ref{table:schedule}.

We plan to make the workshop highly interactive by engaging people in group discussions.
As a workshop is a place where future research is \textit{designed}, we plan to make EmpathiCH a living experiment.
Every interactions (e.g., Q\&A sessions, panel) in presence and virtual (e.g., audio and video feed) will be recorded and later collaboratively analyzed to understand the role of empathy in such context.
The analysis may result in potential publications co-authored by all the organizers and new cross-disciplinary collaborations for future projects.
The follow-up studies will deepen the understanding of empathy and lie the foundation of new guidelines for empathy-centric design.

The workshop will comprehend two main sessions. In the first part of \textbf{presentations} we will follow a traditional conference format.
It will start with a warm-up session consisting in a sensitizing discussion where we will ask the participant to quickly share their thought on the concept of empathy and their expectations for this workshop.
Then we will have a keynote talk from an invited speaker followed by the presentations of the papers accepted to the workshop followed by a moderated discussion.
We will use the \textit{minute madness} format.
Each author will present their work in one minute using one slide, which will be collected before the workshop to ensure a smooth flow.
Finally, we will close the first part with a panel of experts discussing the topic of empathy for multiple stakeholder design in light of the insight gathered so far in the workshop.

Building on the works presented and inspired by the ideas discussed in the first part of the workshop, the second part, \textbf{ideation and mapping}, will employ a participatory format.
At the beginning of the interactive session, organizers and participants will collectively come up with ideas on the Miro board (Eliciting Themes) and cluster these ideas into themes (Cluster Themes).
Then, each theme will be assigned to a specific group of 3-4 participants in the Zoom breakout rooms to develop topics of interest (Ideation Session) and consolidate the topics into executable research agendas (Mapping Session).
There will be various group activities that promote ideation, mapping, and synthesis.
After that, there will be a session of short presentations by each group to discuss each theme.
We expect the outcome to be a list of actionable points or research questions that guide future research of applying empathy in the design process of projects or tools.
The outcome will be placed on our workshop website for public access.

%In the afternoon's interactive sessions specifically, we will develop various strategies to gamify some discussions, for example, by assigning different roles to the participants with varying levels of empathy.

%In this way, we can later assess the impact of empathy on conversations when analyzing the collected research data.

\section{Post-Workshop Plan}

We will arrange a networking event immediately after the workshop to continue the discussion informally.
During and following the workshop, accepted papers, videos, slides, discussion results, and outcomes (i.e., Miro board) will be published on the workshop website. 
Following the workshop, drawing on the workshop submissions and discussions, we will propose a journal special issue or a book in the Springer Series on Human-Computer Interaction.
We will also consolidate and disseminate the result of the workshop in a conference, journal, or magazine article (such as the ACM Interactions), co-authored by all the attendees.
In this way, we create a professional network and encourage participants to collaborate on future ideas, projects, or publications around the developed research agendas.
Finally, we will set up a repository to share and upload acquired research data amongst the attendees, and another one to facilitate collective analysis of this multi-modal data.
\section{Remote and Onsite Plans}\label{section:hybrid-plan}

This workshop will be held in hybrid form, taking into account both the onsite and remote participants.
All participants will receive a Zoom link before the workshop via email.
In the physical workshop room, we will set up a camera that points to the entire room and join the Zoom call with the camera.
We will also set up a microphone for the remote participants to ask questions to the paper presenter directly.
In addition, remote participants can choose to type the questions in the chat, and the workshop organizers can read the questions to the paper presenter.
For interactive activities, we will assign groups and distribute remote participants among them.
Each group will be placed in a breakout room on Zoom, where the onsite participants will join to have conversations with the remote people.
This is to avoid a situation where the remote participants are isolated from the onsite ones.

To promote communication between participants who cannot physically attend the workshop and live in different time-zones, we plan to support asynchronous interactions.
For example, we will make the videos of the accepted works available online before the workshop.
We will also invite people to engage independently with the authors and attendees by tweeting the workshop content on Twitter with a specific hashtag to track online conversations.
Also, to include them in the interactive session (e.g., the warm-up), we will ask them to provide some inputs before the workshop so that we can use them to kick-start the discussion.
Finally, we will publish the content generated during the ideation and mapping session on our website and Twitter so the participants can add their own insights by replying to the tweets.
\section{250 Words Call for Participation}

While empathy has been proven effective in forming relationships between designers and users, it remains an open question about integrating empathy at a large scale in the design process with multiple stakeholders having diverse (even contradicting) perspectives.
In this one-day interactive workshop, we will formulate and develop a set of future challenges and strategies for the ``Empathy-Centric Design at Scale'' research agenda.
We will investigate the role of empathy and the supporting technologies in the design process at scale.
We invite 4-6 page submissions that include position, work-in-progress, provocation, demo, or poster papers in the SIGCHI Full paper format addressing questions such as (but not limited to):
\begin{itemize}
    \item Which facets of empathy can be elicited in which context?
    \item How can we embed empathy in different co-design contexts to lead to inclusive solutions?
    \item How can technologies empower empathy design for various purposes across and within multiple stakeholder groups?
    %\item How can technologies empower creating empathy across and within multiple stakeholder groups? (merged with the previous question)
    %\item What are the co-design methods that can make diverse communities empathize with each other? Andrea: similar to the one above?
    \item How can empathy be employed to reach broader communities and gather unbiased data?
    \item How can we develop or adopt digital tools to elicit empathy at different scales?
\end{itemize}
We aim to assemble a multidisciplinary professional network that involves people in HCI, AI, social science, design, psychology, and health from universities, companies, non-profit organizations, and government sectors.

Submissions should be submitted via Easychair\footnote{\url{https://easychair.org/conferences/?conf=empathich2022}} and will be selected based on novelty, provocativeness, quality, and relevance to the workshop.
Please direct queries to Andrea Mauri (a.mauri@tudelft.nl). Further information is available on the workshop website\footnote{\url{http://www.empathich.io/}}.
At least one author of each accepted paper must attend the workshop, and all participants must register for the workshop for at least one day of the conference.

%\begin{acks}
%To Robert, for the bagels and explaining CMYK and color spaces.
%\end{acks}

\bibliographystyle{ACM-Reference-Format}
\bibliography{sample-base}

%\appendix

\end{document}